\RequirePackage{fix-cm}
\documentclass{svjour3}                     
\smartqed  
\usepackage{graphicx}
\newcommand{\be}{\begin{equation}}
\newcommand{\ee}{\end{equation}}
\def\dif{\mathop{\rm \,d}\nolimits}
\def\ci{\mathop{\textrm{i}}\nolimits}

\begin{document}


\title{Null conformal Killing-Yano tensors and Birkhoff theorem}

\titlerunning{Null Killing-Yano tensors}        

\author{Joan Josep Ferrando        \and
   Juan Antonio S\'aez 
}


\institute{Joan Josep Ferrando \at
              Departament d'Astronomia i Astrof\'{\i}sica, Universitat
de Val\`encia, E-46100 Burjassot, Val\`encia, Spain. \\
Observatori Astron\`omic, Universitat
de Val\`encia, E-46980 Paterna, Val\`encia, Spain. \\
                           \email{joan.ferrando@uv.es}           
           \and
          Juan Antonio S\'aez  \at
               Departament de Matem\`atiques per a l'Economia i l'Empresa,
Universitat de Val\`encia, E-46071 Val\`encia, Spain
}

\date{Received: date / Accepted: date}

\maketitle

\begin{abstract}
We study the space-times admitting a null conformal Killing-Yano tensor whose divergence defines a Killing vector. We analyze the similitudes and differences with the recently studied non null case (Gen. Relativ. Grav. (2015) {\bf 47} 1911). The results by Barnes concerning the Birkhoff theorem for the case of null orbits are analyzed and generalized.

\end{abstract}

\section{Introduction}

In a recent paper \cite{fs-KYB} we have outlined the close relationship between the Jebsen-Birkhoff theorem \cite{Jebsen} \cite{birkhoff} and the results by Tachibana \cite{tachibana} and Hugshton and Sommers \cite{hougshton-sommers} concerning conformal Killing-Yano (CKY) tensors. We have weakened the hypothesis of the generalized version (see \cite{bona} and references therein) of the Jebsen-Birkhoff theorem by showing that the existence of an additional Killing vector does not necessarily require a three-dimensional isometry group on two-dimensional orbits but only the existence of a CKY tensor. Our study analyzes the non null case in detail, and it gives several levels of generalization of the Jebsen-Birkhoff theorem.

In the present paper we offer a similar study for the null case and show that the space-times admitting a null CKY tensor, whose divergence defines a Killing vector, belong to three classes: the generalized pp-waves, a $\Lambda$-generalization of these ones, and a family of conformally flat metrics. Our study allows us to analyze and generalize the Barnes results which extend the Jebsen-Birkhoff theorem to the case of null orbits \cite{barnes} \cite{barnes2}. We also compare the results with the non null case considered in \cite{fs-KYB}.

This paper is organized as follows. In section \ref{sec2} we summarize some concepts and known properties of CKY tensors, and we analyze in detail the integrability conditions when the CKY tensor $A$ is a null two-form and $Z_1 = \delta A$ and $Z_2 = \delta *A$ are Killing vectors.
The non degenerate case $Z_1 \wedge Z_2 \not=0$ is studied in section \ref{bnot0}, and we show that the metric is conformally flat and could be interpreted as a radiation pure deformation of a curvature constant space-time.
In section \ref{sec-b0} we study the case $Z_1 \wedge Z_2 =0$ (when a Killing Yano tensor $A$ exists) which leads to two subcases: if $A$ is a covariantly constant two-form we have the generalized pp-waves \cite{steele} and, otherwise, we obtain the Siklos class of solutions \cite{siklos}, which are pure radiation fields with cosmological constant and  conformal to pp-waves. Section \ref{sec-barnes} is devoted to analyzing the results by Barnes on the space-times admitting a three-parameter isometry group with two-dimensional null orbits. After discussing our results, we will apply them to extend and interpret Barnes issues. 

Here we work on an oriented space-time with a metric tensor
$g$ of signature $\{-,+,+,+\}$ and metric volume element $\eta$. The Riemann, Ricci and Weyl tensors, and the scalar curvature
are defined as given in \cite{kramer} and are denoted, respectively,
by $Riem$, $Ric$, $W$ and $R$. For the metric product of two vectors we
write $(X,Y) = g(X,Y)$. If $A$ and $B$
are 2-tensors, $A \cdot B$ denotes the 2-tensor $(A \cdot
B)^{\alpha}_{\ \beta} = A^{\alpha}_{\ \mu} B^{\mu}_{\ \beta}$, $A^2
= A \cdot A$, $[A,B]$ denotes their commutator $[A,B]= A\cdot B - B \cdot A$, $A(X,Y) = A_{\alpha \beta} X^{\alpha} Y^{\beta}$, $A(X) = A_{\alpha \beta} X^{\beta}$, and
$(A,B) = \frac12 A_{\alpha \beta} B^{\alpha \beta}$. For a vector $X$ and a (p+1)-tensor $t$, $i(X)t$ denotes the
inner product, $[i(X)t]_{\underline{p}} = X^{\alpha}t_{\alpha
\underline{p}}$, the underline denoting multi-index. And if $\omega$ is a (p+1)-form, $\delta \omega $ denotes its exterior codifferential, $(\delta \omega)_{\underline{p}} = -\nabla_{\alpha} \omega^{\alpha}_{\ \underline{p}}$. 

\section{On space-times admitting a null CKY tensor ${\cal A}$} 
\label{sec2}
  
A two-form $A$ is a conformal Killing-Yano tensor if, and only if, its associated self-dual two-form
  ${\cal A} = \frac{1}{\sqrt{2}} (A - \ci *A)$ satisfies the CKY equation, which can be written as \cite{fs-KY}:
\begin{equation} \label{CKYeq}
3 \nabla {\cal A} =  2 i({\cal Z}) {\cal G} \, , \qquad  {\cal Z} \equiv \delta {\cal A}  \, ,
\end{equation}
where $*$ is the Hodge dual operator. The 4-tensor ${\cal G}$ is the endowed metric
on the 3-dimensional complex space of the self-dual two-forms, ${\cal G}=\frac{1}{2}(G-\textrm{i} \; \eta)$, $G$
being the metric on the space of 2--forms, $G_{\alpha \beta \gamma \delta} = g_{\alpha
\gamma} g_{\beta \delta} - g_{\alpha \delta} g_{\beta \gamma}$.

The integrability conditions of the CKY equation (\ref{CKYeq}) lead to constraints on the Petrov-Bel type. In fact the following proposition holds \cite{glass-kress}:
\begin{proposition} \label{propo-weyl}
If a space-time admits a non null CKY tensor ${\cal A}$, then the Weyl tensor is type O or type D, and in type D case ${\cal A}$ is an eigen-two-form of the Weyl tensor. 

If a space-time admits a null CKY tensor ${\cal A}$, then the Weyl tensor is type O or type N, and in type N case ${\cal A}$ is an eigen-two-form of the Weyl tensor.
\end{proposition}

Moreover, the integrability conditions of the CKY equation also constrain the Ricci tensor. More precisely, we have \cite{tachibana} \cite{fs-KYB}:
\begin{proposition} \label{propo-ricci}
If a space-time admits a conformal Killing-Yano tensor ${\cal A}$, then ${\cal Z} \equiv \delta {\cal A}$
is a (complex) Killing vector (or it vanishes) if, and only if, $[ {\cal A} , Ric]=0$.\\
\end{proposition}
Finally, a standard tensorial calculation leads to the following constraint which completes the integrability conditions to the CKY equation: 
\begin{equation} \label{dZ}
\dif {\cal Z} - \ci *\dif {\cal Z} = \frac{R}{2} {\cal A} - 3 {\cal W}({\cal A}) \, .
\end{equation}

\subsection{The case when ${\cal Z} \equiv \delta {\cal A}$ is a Killing vector}
\label{sec3}

A null CKY tensor  ${\cal A}$ can be written as ${\cal A} = \ell \wedge m$, where $\ell$ is its (null) fundamental vector, and $m$ is a complex null vector satisfying $(m, \ell)=0, (m, \bar{m})=1$, and fixed up to change $m+ c \ell$. 

Proposition \ref{propo-weyl} implies that the self-dual Weyl tensor, ${\cal W}=\frac{1}{2}(W-\textrm{i} \; * W)$, takes the expression:
\begin{equation} \label{weyl}
{\cal W} = \psi {\cal A} \otimes {\cal A}\, ,
\end{equation}
where the complex function $\psi$ vanishes if the space-time is type O. Otherwise the space-time is type N.

On the other hand, if ${\cal Z} \equiv \delta {\cal A}$ is a (complex) Killing vector, proposition \ref{propo-ricci} implies that the Ricci tensor is of Segr\`e types $[(211)]$ or $[(1111)]$, that is:
\begin{equation}  \label{Ricci}
Ric = \sigma \ell \otimes \ell + \Lambda g \, \qquad R = 4 \Lambda \, .
\end{equation}
Moreover, equation (\ref{dZ}) becomes:
\begin{equation}  \label{nablaZ}
\nabla {\cal Z} = \frac12 \dif {\cal Z} = \frac{\Lambda}{2} {\cal A} + \bar{\cal H}  \, , \qquad {\cal G}(\bar{\cal H}) = 0  \, .
\end{equation} 

We have ${\cal A}^2 = 0$. Then, from (\ref{CKYeq}) we obtain $(\nabla{\cal A}, {\cal A} )=0$, which leads to ${\cal A}({\cal Z}) =0$ and $i({\cal Z})\nabla{\cal A}=0$. On the other hand, from (\ref{nablaZ}), and considering that every self-dual two-form commutes with every anti-self-dual one, we have $[\nabla {\cal Z}, {\cal A}]=0$. This condition together with $i({\cal Z})\nabla{\cal A}=0$ imply ${\cal L}_{\cal Z} {\cal A}=0$. So we get:
\begin{equation} \label{LZA}
 {\cal A}({\cal Z})=0 \, , \qquad  i({\cal Z}) \nabla {\cal A} = 0 \, , \qquad  {\cal L}_{\cal Z} {\cal A} = 0 \, , \qquad [\nabla {\cal Z}, {\cal A}]=0 \, .
\end{equation}
And then, the first equation in (\ref{LZA}) implies:
\begin{equation}  \label{Z-ell}
{\cal Z} = a \ell - b m \, , \quad ({\cal Z}, {\cal Z}) = 0 \, , \quad {\cal Z} \wedge \ell  = b {\cal A}  = {\cal G} ({\cal Z} \wedge \ell)   \, , \quad \bar{\cal G}({\cal Z} \wedge \ell)=0 \, .
\end{equation}

Now, from the integrability condition of the Killing equation,
\begin{equation} \label{integraKE}
\nabla \nabla {\cal Z} = i({\cal Z}) Riem \, ,
\end{equation}
and taking into account (\ref{weyl}), (\ref{Ricci}), (\ref{nablaZ}), (\ref{LZA}), (\ref{Z-ell}), we obtain:
\begin{equation} \label{integraKE-b}
\nabla \bar{\cal H} = \frac{\Lambda}{3} i({\cal Z}) \bar{\cal G} + i({\cal Z})\bar{\cal W}  \, ,\qquad \dif \Lambda   = \sigma \, b \,  \ell \, .
\end{equation}

On the other hand, $({\cal Z}, {\cal Z}) = 0$ implies $i({\cal Z})\nabla{\cal Z} =0$, and with (\ref{nablaZ}) and (\ref{LZA}) lead to:
\begin{equation}  \label{F-Z}
\bar{\cal H}({\cal Z})=0 \, .
\end{equation}
If we derive (\ref{F-Z}) and take into account (\ref{nablaZ}) and (\ref{integraKE-b}), we obtain:
\begin{equation}  \label{F-F}
\bar{\cal H}^2 + \frac{\Lambda}{2}{\cal A} \cdot \bar{\cal H} - \frac{\Lambda}{6}{\cal Z} \otimes {\cal Z} + \bar{\psi}\bar{\cal A}({\cal Z}) \otimes \bar{\cal A} ({\cal Z})=0 \, .
\end{equation}
Now if we calculate the trace of this last equation and we consider that ${\cal A}$ is a null 2-form, $({\cal A}, {\cal A}) = (\bar{\cal A}, \bar{\cal A}) =0$, we arrive to $(\bar{\cal H}, \bar{\cal H}) =0$, and then $\bar{\cal H}^2 =0$. Moreover, $\bar{\cal A} ({\cal Z})= - b \ell$. Consequently (\ref{F-F}) becomes:
\begin{equation}  \label{F-A}
\Lambda ( 3 {\cal A} \cdot \bar{\cal H} - {\cal Z} \otimes {\cal Z}) + 6 \bar{\psi} \, b ^2 \, \ell \otimes \ell= 0 \, .
\end{equation}

From now on we consider two cases which lead to two different families of metrics. 

\section{Solutions with ${\cal Z} \wedge \bar{\cal Z} \not=0$}
\label{bnot0}

This condition is equivalent to $b \not= 0$, and then we can introduce a new $m$ by the change $m - \frac{a}{b} \ell \ \rightarrow \ m$. With this choice we have ${\cal Z} = - b m$, and a unique null base $\{\ell, k, m, \bar{m}\}$ is determined. Then (\ref{F-Z}) implies $\bar{\cal H}(m)=0$ and, consequently, $\bar{\cal H} = \beta k \wedge m$. Now (\ref{F-A}) states:
\begin{equation}  \label{psi0}
\Lambda ( 3 \beta - b^2) = 0 \, , \qquad  \psi = 0 \, .  \
\end{equation}
Thus, the space-time is conformally flat. Moreover, if $\sigma \not=0$, (\ref{integraKE-b}) implies $\dif \Lambda\not=0$ and $b= \bar{b}$, and then $\beta$ is a real scalar, $3 \beta = b^2$. In this case (\ref{nablaZ}) becomes
\begin{equation}  \label{nablaZ-b}
\nabla {\cal Z} = \left(\frac{\Lambda}{2} \ell + \frac{b^2}{3} k \right) \wedge m \, , \qquad  {\cal Z} = -b m \, ,
\end{equation} 
and consequently, $i(\bar{\cal Z}) \nabla {\cal Z}$ is a real vector, that is $ [{\cal Z}, \bar{\cal Z}]=0$. On the other hand, from (\ref{LZA}) we have ${\cal L}_{\cal Z} {\cal A} = 0$, and (\ref{CKYeq}) and (\ref{nablaZ-b}) imply ${\cal L}_{\bar{\cal Z}} {\cal A} = 0$. Thus, if we consider the real Killing vectors $Z_i$ defined by $\sqrt{2} {\cal Z} = Z_1 + \ci Z_2$, we can state: 

\begin{proposition} \label{propo-b}
Let us consider a non constant curvature space-time admitting a null conformal Killing-Yano tensor $A$ such that $Z_1 \equiv \delta A$ and $Z_2 \equiv \delta \! * \! A $ are Killing vectors verifying $Z_1 \wedge Z_2 \not= 0$. Then: (i) $\{Z_1, Z_2\}$ define a commutative algebra: $[Z_1,Z_2]=0$, (ii) the CKY tensor is $Z_i$-invariant: ${\cal L}_{Z_1} A = {\cal L}_{Z_2} A = 0$, (iii) the space-time is conformally flat ($W=0$), and (iv) the Ricci tensor writes $Ric = \sigma \ell \otimes \ell + \Lambda g$, with $\dif \Lambda \not=0$, $\ell$ being the fundamental vector of $A$.
\end{proposition}

In order to study the metric line element of the space-times considered in proposition \ref{propo-b}, we can expand equations  (\ref{CKYeq}),  (\ref{integraKE-b}) and (\ref{nablaZ-b}) by taking into account that ${\cal A}= \ell \wedge m$,  $\bar{\cal H} = \frac{1}{3} b^2 k \wedge m$. Then, after a standard tensorial calculation, we obtain the following exterior system for the null co-base $\{\ell, k, m, \bar{m}\}$:
\begin{eqnarray} \label{exterior-b}
\dif \ell = - \frac23 b \, \ell \wedge k \, , \qquad \dif k = 0 \, , \qquad \dif m = \dif \ln b \wedge m \, ,\\
\dif b = - \frac{\Lambda}{2} \ell - \frac13 b^2 k \, , \qquad \quad \qquad \dif \Lambda = \sigma \, b \, \ell \, .
\end{eqnarray}
From this system we have that a coordinate system $\{u,v,\xi,\bar{\xi}\}$ exists such that:
\begin{equation} \label{coordinates-b}
\ell = \alpha \dif u \, , \qquad k = \dif v \, , \qquad m = \frac{b}{3} \dif \xi \, ,
\end{equation}
where $\alpha = \alpha(u,v)$, $b=b(u,v)$ are submitted to the equations:
\begin{equation}
\frac{\dot{\alpha}}{\alpha} = - \frac23 b \, , \quad b'= -\frac12 \alpha \Lambda \, , \quad \dot{b} = - \frac13 b^2 \, , \quad \Lambda = \Lambda(u) \, ,
\end{equation}
and where prime and dot denote differentiation with respect to $u$ and $v$ respectively. We can easily integrate these equations and we obtain:
\begin{equation}
b = \frac{3}{v+f(u)}  \, , \quad \alpha = \frac{h(u)}{[v+f(u)]^2} \, , \quad h(u) \equiv \frac{6 f'(u)}{\Lambda(u)}  \, .
\end{equation}
Now we can introduce a new coordinate $\tilde{u}$ defined by $\dif \tilde{u} = h(u) \dif{u}$. Then, again denoting $\tilde{u}$ as $u$, we have $\alpha = [v+f(u)]^{-2}$ and $\Lambda = 6 f'(u)$, and we arrive to the following.
\begin{theorem} \label{theo-b-metrica}
For the space-time admitting a null conformal Killing-Yano tensor $A$ such that $Z_1 \equiv \delta A$ and $Z_2 \equiv \delta \! * \! A $ are Killing vectors verifying $Z_1 \wedge Z_2 \not= 0$, the metric line element is given by
\begin{equation} \label{metric-b}
\dif s^2 = \frac{1}{[v+f(u)]^2}[- 2\dif u  \dif v + 2 \dif \xi  \dif \bar{\xi}] \, .
\end{equation}
And the Ricci tensor takes the expression:
\begin{equation} \label{ricci-b}
Ric = \sigma \ell \otimes \ell + \Lambda g \, , \qquad \sigma =2 f''(u)[v + f(u)]^3 \, , \quad \Lambda = 6 f'(u) \, .
\end{equation}
\end{theorem}
Note that the statement of this theorem includes the metrics of constant curvature which correspond to $\dif \Lambda = 0$, $\sigma=0$, and $f(u) = \frac{\Lambda}{6} u$. This case has been removed in proposition \ref{propo-b} and in obtaining the metric line element (\ref{metric-b}). Nevertheless, a simple calculation shows that the Ricci tensor of the conformal metric (\ref{metric-b}) is also given by (\ref{ricci-b}) in this case. 
\
\\[2mm]
{\it Remark 1} In the coordinate system $\{u, v, \xi, \bar{\xi}\}$ the metric $g$ takes the conformally flat form (\ref{metric-b}), $g = \phi^2 \tilde{g}$, where $\phi = [v + f(u)]^{-1}$ and $\tilde{g}$ is the flat metric. Moreover ${\cal A} = \phi^3 \dif u \wedge \dif \xi$ is a CKY tensor, and $\tilde{\cal A} = \dif u \wedge \dif \xi$ is a CKY tensor of the flat metric, in accordance with the known invariance of the CKY equation by a conformal transformation \cite{jezierski}.
\
\\[2mm]
{\it Remark 2} Note that metrics (\ref{metric-b}) may be interpreted as a deformation of the space-times of constant curvature. When the function $f''(u)$ does not vanish, we have a pure radiation solution with a $\Lambda$-term depending on the null coordinate $u$, $\Lambda = 6 f'(u)$. The radiation term in the energy tensor,  $T_r = \sigma \ell \otimes \ell$, never corresponds to a Maxwell solution. Indeed, the Maxwell null fields with fundamental direction $\ell$ are ${\cal F} = \zeta(u, \xi) \phi^{-3} {\cal A}$, with an associated electromagnetic energy $T_m= \zeta \bar{\zeta} \phi^{-6} \ell \otimes \ell$, which is not compatible with the expression (\ref{ricci-b}) of $\sigma$.
\
\\[2mm]
{\it Remark 3} In the non null case studied in \cite{fs-KYB}, condition $Z_1 \wedge Z_2 \not= 0$ leads to a family of metrics (of Petrov-Bel type D or type O) containing the generalized C-metrics. A more detailed analysis shows that the conformally flat case is not compatible in this regular case. Nevertheless, in the null case studied here, only the conformally flat case is possible. Even so, proposition \ref{propo-b} shows the great similarities between the two cases. Indeed, properties (i) and (ii) hold in both the null and non null cases. Moreover, when $A$ is a simple non null CKY tensor, $Z_i$ are hypersurface orthogonal Killing vectors \cite{fs-KYB}, and here $A$ is null, and consequently simple, and $Z_i$ are also hypersurface orthogonal Killing vectors. This fact is a direct consequence of (\ref{nablaZ-b}).

\section{Solutions with ${\cal Z} \wedge \bar{\cal Z} =0$}
\label{sec-b0}

Now we have $b=0$, and ${\cal Z} = a \, \ell$ . Then (\ref{F-Z}) implies $\bar{\cal H}(l)=0$ and, consequently, $\bar{\cal H} = \pi \bar{\cal A}$. From these relations, (\ref{F-A}) states:
\begin{equation}  \label{b0}
\Lambda ( 3 \pi + a^2) = 0  \, .  \
\end{equation}
Moreover, the second equation in (\ref{integraKE-b}) implies $\dif \Lambda = 0$.

We have ${\cal Z} =\frac{1}{\sqrt{2}} (Z_1 + \ci Z_2) = a \ell$. Then, the real Killing vectors $Z_1, Z_2$ are parallel and, consequently, they differ in a constant. Thus $a$ has a constant argument $\theta$. If we make a constant duality rotation, ${\cal F} =e^{- \ci  \theta}{\cal A}$ is a solution of the CKY equation. Moreover the real two form $*F$ is a Killing-Yano tensor, and we have the Killing vector $Z \equiv \delta F = \sqrt{2} a \ell$, $a$ being a real scalar. From now on, in this section we will work with the CKY tensor ${\cal F}$. Now (\ref{nablaZ}) and $\bar{\cal H} = \pi \bar{\cal F}$ lead to $\pi = \frac{\Lambda}{2}$, and then (\ref{nablaZ}) writes:
\begin{equation}  \label{nablaZ-0}
\nabla Z =  \frac{\Lambda}{\sqrt{2}}[ {\cal F} + \bar{\cal F}]  = \Lambda \, F =  \Lambda \, \ell \wedge p  \, , \qquad m = \frac{1}{\sqrt{2}} (p + \ci q) \, .
\end{equation}

When $\Lambda = 0$, (\ref{nablaZ-0}) implies $\nabla Z =0$. Then, from $Z \equiv \delta F = \sqrt{2} a \ell$ we obtain that ${\cal F}$ is recurrent, $\nabla {\cal F} = v \otimes {\cal F}$, a condition which is incompatible with the CKY equation unless $Z=0$. Consequently,  the CKY tensor ${\cal F}$ is covariantly constant, $\nabla {\cal F}=0$, and the space-time is a generalized pp-wave \cite{steele}, $\ell$ being the covariant constant null vector, $\nabla \ell = 0$. Then we arrive at the following result:
\begin{proposition} \label{prop-b00-metrica}
Let us consider the space-times admitting a null conformal Killing-Yano tensor $A$ such that $Z_1 \equiv \delta A$ and $Z_2 \equiv \delta \! * \! A $ are Killing vectors verifying $Z_1 \wedge Z_2= 0$. Then $\Lambda =0$ if, and only if, $Z_i=0$, and then the space-time is a pp-wave, whose metric line element is given by
\begin{equation} \label{metric-pp}
\dif s^2 = - 2\dif u  \dif v - 2 H \dif u^2 + 2 \dif \xi  \dif \bar{\xi} \, , \quad  H=H(u, \xi, \bar{\xi}) \, .
\end{equation}
The covariantly constant two-form is:
\begin{equation}
{\cal A} = \ell \wedge m \, , \quad   \ell = \dif u \, , \quad m = \dif \xi \, ,
\end{equation}
and the Ricci and Weyl tensors take the expressions:
\begin{equation} \label{ricci-b0}
Ric = \sigma \ell \otimes \ell  \, , \quad \sigma = H_{\xi \bar{\xi}} \, ; \qquad {\cal W} = \psi {\cal A} \otimes {\cal A} \, ,   \quad  \psi = H_{\xi \xi}   \, .
\end{equation}
\end{proposition}

On the contrary, when $\Lambda \not=0$, (\ref{b0}) implies $a^2=- 3 \pi = - \frac32 \Lambda =\  constant\ > 0$. So, $\ell$ is a Killing vector and 
\begin{equation}  \label{nabla-l}
\nabla \ell = \gamma  \ell \wedge p \, \qquad   \gamma \equiv  \sqrt{- \frac{\Lambda}{3}}  \, .
\end{equation}
From this expression we obtain:
\begin{equation}
\ell = \omega^{-2} \dif u \, , \qquad  p =  \frac{1}{\gamma} \dif \ln \omega \, .
\end{equation}
Moreover, the integrability conditions of (\ref{nabla-l}) lead to $\nabla p = \gamma[p \otimes p - g + \mu \ell \otimes \ell]$. Then, if we consider the conformal metric $\tilde{g} = \omega^{2} g$, and $\tilde{\nabla}$ denotes its associated Levi-Civita connection, we obtain: 
\begin{equation}
\tilde{\nabla} \tilde{\ell} = 0 \, , \quad \tilde{\ell} = \omega^2 \ell \, ; \qquad \tilde{\nabla} \tilde{p} = \gamma \mu \omega \ell \otimes \ell \, , \quad \tilde{p} = \omega p \, .
\end{equation}
Consequently, $\tilde{g}$ is a pp-wave and the covariantly constant two-form is $\tilde{F} =  \tilde{\ell} \wedge \tilde{p}$. From these results, proposition \ref{prop-b00-metrica}, and considering the changes in the Ricci and Weyl tensors by a conformal transformation, we obtain:  
\begin{theorem} \label{theo-b0-metrica}
For the space-times admitting a null conformal Killing-Yano tensor $A$ such that $Z_1 \equiv \delta A$ and $Z_2 \equiv \delta \! * \! A $ are Killing vectors verifying $Z_1 \wedge Z_2= 0$, the metric line element is given by
\begin{equation} \label{metric-b0}
\dif s^2 = \omega^{-2}[- 2\dif u  \dif v - 2 H \dif u^2 + \dif x^2 +  \dif y^2] , \   H=H(u, x, y) , \  \omega = 1+\gamma x  .
\end{equation}
Moreover the metric admits the Killing-Yano tensor $*F$, where:
\begin{equation}
F = \ell \wedge p = \omega^{-3} \dif u \wedge \dif x \, ,
\end{equation}
and the Ricci and Weyl tensors take the expressions:
\begin{equation} \label{ricci-weyl-b0}
\begin{array}{l}
Ric = \sigma \ell \otimes \ell + \Lambda g \, , \quad \sigma = \omega^4[ H_{xx} + H_{yy}] - 2 \gamma \omega^3 H_x \, , \quad \Lambda = -3 \gamma^2 \, ; \\[2mm] 
{\cal W} = \psi {\cal F} \otimes {\cal F} \, ,   \quad \qquad   \psi = \omega^4[ H_{xx} - H_{yy} - 2 \ci  H_{xy}]   \, .
\end{array}
\end{equation}
\end{theorem} 
The expression of $\omega=\omega(x)$ follows from the integration of equation $\dif x = \tilde{p} =\omega p = \gamma^{-1} \dif \omega$.
\\[2mm]
{\it Remark 4} Note that this theorem includes proposition \ref{prop-b00-metrica} as the particular case $\gamma=0$. We see that the family of metrics (\ref{metric-b0}) are a counterpart with cosmological constant of the generalized pp-waves, and only a $\Lambda < 0$ is possible. These $\Lambda$-{\it pp-waves} (\ref{metric-b0}) were considered by Siklos \cite{siklos} (see also \cite{podolsky}) and they could have been obtained here by integrating an exterior system as we have done in the previous case studied in section \ref{bnot0}. Nevertheless, we have chosen to use previously known results and the fact that the metrics is conformal to a pp-wave. Note that they are plane-fronted waves (Kundt's class \cite{kramer}).

On the other hand, in order to compare with the non null case it is worth stating the following.
\begin{proposition} \label{propo-b0}
In a non constant curvature space-time admitting a null conformal Killing-Yano tensor $A$ such
that $[A , Ric]=[*A, Ric]=0$, let us consider $Z_1 \equiv \delta A$ and $Z_2 \equiv \delta \! * \! A $.
Then, we have the following equivalent conditions:
(i) $Z_1 \wedge Z_2 =0$,
(ii) The fundamental vector $\ell$ of the two-form $A$ is a Killing vector (moreover, $\ell \wedge Z_i=0$),
(iii) A constant duality rotation $\theta$ exists such that if $F = \sin \theta A - \cos \theta *A$, then $*F$ is a Killing-Yano tensor,
(iv) $K = F^2$ is a Killing tensor,
(v) The Killing two-form $\dif Z$ of the Killing vector $Z = \delta F$ is aligned with $F$: $[F, \dif Z] = 0$.
\end{proposition}
Equation (\ref{nablaZ-0}) and the reasoning above prove that (i) implies (ii), (iii) and (v). Moreover (ii) and (iv) are trivially equivalent because $K = F^2 = - \ell \otimes \ell$, and it is known that (iii) implies (iv). On the other hand, if (i) does not hold, neither (ii) nor (v) holds as a consequence of the study of the case $Z_1 \wedge Z_2 \not=0$ presented in section \ref{bnot0}. 
\\[2mm]
{\it Remark 5} In the non null case studied in \cite{fs-KYB} condition $Z_1 \wedge Z_2 =0$ leads to a wide family of metrics which contains the Kerr-NUT charged and vacuum solutions and their counterpart with cosmological constant. In the null case studied here we obtain the pp-waves and their $\Lambda$-counterpart. But proposition \ref{propo-b0} shows that both cases present similar properties. Moreover, here $A$ is simple and $Z$ is a hypersurface orthogonal Killing vector, similarly to what happens when $A$ is simple in the non null case. In this case we also have the invariance ${\cal L}_Z A =0$.   

\section{Analysis and extension of the Barnes results}
\label{sec-barnes}

The extension of the Jebsen-Birkhoff theorem to the null orbits was studied by Barnes in two papers \cite{barnes} \cite{barnes2}. Below, we summarize some of these Barnes results in order to analyze and generalize them by using our results presented in this paper.
\\[2mm]
{\it Barnes result 1.-} \cite{barnes2} The metric line element of the space-times admitting a three parameter isometry group with two-dimensional null orbits ($G_3(2n)$)  is given by:
\begin{equation} \label{Barnes-metric}
\dif s^2 =  \nu^2(X,U)[- 2 \dif V \dif U + 2 \epsilon(X,U) \dif U^2 + \dif Y^2] + \dif X^2  \, .
\end{equation}
{\it Barnes result 2.-} \cite{barnes} \cite{barnes2} A space-time which admits a $G_3(2n)$ and which is a pure radiation field, $Ric = \sigma \ell \otimes \ell$, is a (generalized) pp-wave. Its metric line element is given by (\ref{Barnes-metric}) with $\nu=\nu(U)$. Moreover, if the space-time is a vacuum or an invariant Einstein-Maxwell field, it is a plane wave space-time.
\\[2mm]
{\it Barnes result 3.-} \cite{barnes2} A space-time which admits a $G_3(2n)$ and which is a vacuum solution with cosmological constant, $Ric = \Lambda g$, is a plane-fronted wave with a metric line element  given by (\ref{Barnes-metric}), with $\nu = \frac{1}{U}e^{\gamma X}$, $\epsilon = \epsilon_0(U)e^{-3 \gamma X}$, $\Lambda = -3 \gamma^2$.
\\[-2mm]

A straightforward calculation leads to the following lemma which is useful to study the relationship between the results by Barnes and ours:
\begin{lemma}  \label{lemma-KY}
The space-times with a $G_3(2n)$ given in (\ref{Barnes-metric}) admit the null Killing-Yano tensor $*F$, where $F= \ell \wedge p$, $\ell = \nu^2 \dif U$, $p = \dif X$.
\end{lemma}
Now we present several consequences of previous sections showing that the existence of a $G_3(2n)$ is a hypothesis of the Barnes results that can be weakened by only imposing the existence of a null Killing-Yano tensor.

Hypothesis of proposition \ref{prop-b00-metrica} equivalently states that a null Killing-Yano tensor exists. On the other hand, a generalized pp-wave is a generalized plane wave when $H$ is quadratic in $\xi, \bar{\xi}$, which can be characterized by $\psi = \psi(u)$ \cite{steele}. If we write this last condition in terms of contractions of $\dif \psi$ with the Killing-Yano tensor $A$, we obtain the following. 
\begin{corollary} \label{cor-ppwaves}
The space-times which admit a null Killing-Yano tensor $A$ and which are pure radiation fields, $Ric = \sigma \ell \otimes \ell$, are the (generalized) pp-waves. Moreover, we have ${\cal W} = \psi {\cal A}\otimes {\cal A}$, and if $A(\dif \psi) = *A(\dif \psi)=0$, the space-time is a plane wave.
\end{corollary}
{\it Remark 6}  This corollary generalizes the Barnes result 2 above by changing the $G_3(2n)$ condition to the weaker one of a null Killing-Yano tensor. This change is similar to that made in our generalization of the Jebsen-Birkhoff theorem for the case of non null orbits \cite{fs-KYB}. Nevertheless, it is worth remarking two important differences with the non null case. On one hand, in the case of non null orbits the generalized Jebsen-Birkhoff theorem implies the existence of a new Killing vector and a $G_4$ exists, and in our extension presented in \cite{fs-KYB} the existence of a Killing-Yano tensor (and no symmetries) implies the existence of a Killing vector. But in the null case, the Killing vector $\ell$ of the pp-waves in the Barnes result 2 was previously a Killing vector of the isometry group $G_3(2n)$. In our extension of the null case, with no Killing vectors {\it a priory}, we have a real Jebsen-Birkhoff-like theorem because a new Killing vector $\ell$ exists. On the other hand, the algebraic type of the Ricci tensor does not change by adding a term $\Lambda g$, which is considered in the usual generalization of the Jebsen-Birkhoff theorem for non null orbits \cite{bona}, but is not considered by Barnes. Below, we extend our results to this case too. Moreover, note that our statement is a necessary and sufficient condition, that is, the pp-waves can be characterized as the pure radiation or vacuum solutions admitting a null Killing-Yano tensor. The pp-waves with a $G_3(2n)$ correspond to $H = B(u,x) + y^2C(u)$, which lead to the metrics considered in the Barnes result 2 with the change $\dif u = \nu^2(U) \dif U$, $v=V+ \frac{\nu'}{2 \nu} Y^2 $, $y = \nu Y$.
\\[2mm]
{\it Remark 7} We can consider the second statement in corollary \ref{cor-ppwaves} as a generalization of the second statement of the Barnes result 2. Now we have no isometry group in the hypothesis, and we have expressed conditions which characterize the full set of the generalized plane waves \cite{steele} in terms of the element which appears in the hypothesis of our statement: the Killing-Yano tensor $A$. Equation $(\dif \psi, \dif \bar{\psi}) =0$ is another equivalent condition, also avoiding the use of coordinates, which turns into plane waves.

Now we consider the more generic compatible Ricci tensor. From theorem \ref{theo-b0-metrica} we can obtain a result that generalizes the first statement in corollary \ref{cor-ppwaves}. Moreover, if we name $\Lambda$-{\it plane waves} the metrics of the family (\ref{metric-b0}) such that $H$ is quadratic in $\xi, \bar{\xi}$, a straightforward calculation leads to the following.
\begin{corollary} \label{cor-lambda-ppwaves}
The space-times which admit a null Killing-Yano tensor $A$ and with a Ricci of Segr\'e types $[(211)]$ or $[(1111)]$, $Ric = \sigma \ell \otimes \ell + \Lambda g$, are the $\Lambda$-pp-waves given in theorem \ref{theo-b0-metrica}. Moreover, we have ${\cal W} = \psi {\cal A} \otimes {\cal A}$, $\psi=\psi_1 + \ci \psi_2$, and the space-time is a $\Lambda$-plane wave if, and only if, $*A(\dif \psi) =0$ and one of the two following conditions holds: (i) $\psi_2 \not=0$, $A(\dif (\psi_1/ \psi_2)) =0$, $(\dif \psi_2)^2 =16 \gamma^2 \psi_2^2$,  (ii) $\psi_2 =0$, $(\dif \psi_1)^2 =16 \gamma^2 \psi_1^2$, $\Delta \psi_1 = 4 \gamma  \psi_1$. 
\end{corollary}
{\it Remark 8} As commented in remark 4, the $\Lambda$-pp-waves in this corollary are the Siklos class of metrics \cite{siklos}, which coincide with a subclass of the plane-fronted waves considered by Ozsv\'ath {\it et al.} \cite{ozsvath}. Siklos studied the admitted isommetry groups depending on specific expressions of function $H$ \cite{siklos}. It is known that the plane waves admit a $G_5$ group of isometries \cite{kramer}, and from the Siklos paper it clearly follows that the $\Lambda$-plane waves can admit a $G_r$ group, $r$ from $1$ to $5$.

As a direct consequence of corollary \ref{cor-lambda-ppwaves}, we obtain.
\begin{corollary} \label{cor-lambda}
The space-times that admit a null Killing-Yano tensor $A$ and are vacuum solutions with cosmological constant, $Ric = \Lambda g$, are plane-fronted waves with a metric line element given by (\ref{metric-b0}), with $H(u,x,y)$ a solution of the equation:
\begin{equation} \label{vacuum-Lambda}
(1+\gamma x)(H_{xx} + H_{yy}) = 2 \gamma  H_x \, .
\end{equation}
\end{corollary}
{\it Remark 8} This corollary extends the Barnes result 3 by changing the $G_3(2n)$ condition to the weaker one of a null Killing-Yano tensor. The Barnes metrics correspond with $H=B(u)\omega^3(x)$ (which is a particular solution to (\ref{vacuum-Lambda})) and the coordinate change $u=-1/U$, $v= V- \frac{Y^2}{2U}$, $\omega(x) = e^{-\gamma X}$, $y= \frac{Y}{U}$.  Moreover, the metrics in corollary \ref{cor-lambda} are the Siklos vacuum solution (with cosmological constant) \cite{siklos}, which has been interpreted as a gravitational wave in the anti-de Sitter universe \cite{podolsky}. In his paper, Siklos gives the explicit expression for the solution of equation (\ref{vacuum-Lambda}).

Finally, we present an alternative statement of theorem \ref{theo-b-metrica}:
\begin{corollary} \label{cor-b}
The space-times which admit a null CKY tensor $A$ (which is not a Killing-Yano tensor) and with a Ricci of Segr\'e types $[(211)]$ or $[(1111)]$, $Ric = \sigma \ell \otimes \ell + \Lambda g$, are the space-times given in theorem \ref{theo-b-metrica}, which admit a commutative group $G_2$. 
\end{corollary}
%
%
{\it Ending remark} In this paper we have studied the space-times admitting a CKY tensor whose divergence determines a Killing vector. For each of the derived families of solutions, we have obtained the metric line element (theorems \ref{theo-b-metrica} and \ref{theo-b0-metrica} and proposition \ref{prop-b00-metrica}), and we have compared them with previously known solutions. Our results extend and clarify the Barnes results concerning the generalization of the Jebsen-Birkhoff theorem for null orbits. By considering a suitable Ricci tensor and without imposing symmetries, corollaries \ref{cor-ppwaves} and \ref{cor-lambda-ppwaves} perform two levels in the generalizations of the Barnes results, and they show that the existence of a Killing-Yano tensor implies that a Killing vector exists. Furthermore, when only a CKY tensor is present, then two Killing vectors exist (corollary \ref{cor-b}). Note the close parallelism between these results and those presented in \cite{fs-KYB} for the non null case. Finally, corollary \ref{cor-lambda} shows that the $\Lambda$-vacuum solution studied by Barnes in \cite{barnes2} is a particular case of the Siklos solution \cite{siklos}.



\begin{acknowledgements}
 This work has been supported by the Spanish ``Ministerio de
Econom\'{\i}a y Competitividad", MICINN-FEDER project FIS2012-33582.
\end{acknowledgements}

\end{document}